\documentclass[apj]{emulateapj}

\usepackage{times}

\shorttitle{Plasma Diagnostics and EIS}
\shortauthors{Warren, Ugarte-Urra, \& Landi}

\begin{document}

%% ------------------------------------------------------------------------------------------
%% --- TITLE PAGE ---------------------------------------------------------------------------
%% ------------------------------------------------------------------------------------------

\title{The Absolute Calibration of the EUV Imaging Spectrometer on \textit{Hinode}}

\author{Harry P. Warren\altaffilmark{1}, Ignacio Ugarte-Urra\altaffilmark{2}, and Enrico
  Landi\altaffilmark{3}}

\affiliation{\altaffilmark{1}Space Science Division, Naval Research Laboratory, Washington, DC
  20375 USA}
\affiliation{\altaffilmark{2}College of Science, George Mason University, 4400 University Drive,
  Fairfax, VA 22030 USA}
\affiliation{\altaffilmark{3}Department of Atmospheric, Oceanic and Space Sciences, University of
  Michigan, Ann Arbor, MI 48109, USA}

%% ------------------------------------------------------------------------------------------
%% --- ABSTRACT -----------------------------------------------------------------------------
%% ------------------------------------------------------------------------------------------

\begin{abstract}
  We investigate the absolute calibration of the EUV Imaging Spectrometer (EIS) on \textit{Hinode}
  by comparing EIS full-disk mosaics with irradiance observations from the EUV Variability
  Experiment (EVE) on the \textit{Solar Dynamics Observatory}. We also use ultra-deep ($>10^5$\,s)
  exposures of the quiet corona above the limb combined with a simple differential emission
  measure model to establish new effective area curves that incorporate information from the most
  recent atomic physics calculations. We find that changes to the EIS instrument sensitivity are
  a complex function of both time and wavelength. We find that the sensitivity is decaying
  exponentially with time and that the decay constants vary with wavelength. The EIS short
  wavelength channel shows significantly longer decay times than the long wavelength channel.
\end{abstract}


%% ------------------------------------------------------------------------------------------
%% --- BODY ---------------------------------------------------------------------------------
%% ------------------------------------------------------------------------------------------

\section{introduction}

Understanding the physical properties of the solar corona is largely dependent on plasma
diagnostics derived from spectroscopic observations. The utility of most plasma diagnostics is
directly related to our understanding of both the underlying atomic physics and the radiometric
calibration of the instrument used to observe them. Unfortunately, establishing and maintaining
the calibration of satellite borne instruments has proven to be challenging, particularly at
extreme ultraviolet (EUV, 50--1200\,\AA) wavelengths where many spectroscopically interesting
emission lines lie.

In this paper we investigate the interplay between the absolute calibration of the EUV Imaging
Spectrometer (EIS \citealt{culhane2007}) on the \textit{Hinode} mission and our ability to measure
the properties of the solar corona. Information on the absolute calibration can be summarized in
the ``effective area,'' which combines the geometrical size and efficiency of the various
components along the optical path of an instrument. EIS observes in two EUV wavelength ranges
(171--212\,\AA\ and 245--291\,\AA) which contain many emission lines that can be used to determine
the temperature and density structure of the corona. Such measurements are critical for
constraining theories of coronal heating. The observation of temperature and density insensitive
emission lines can also be used to evaluate the relative calibration of EIS. As we describe here,
detailed analysis shows that the pre-flight calibration is inconsistent with a number of
established line ratios. These line ratios also indicate that the relative calibration has been
changing with time.

The EIS pre-flight calibration and its evolution early in the mission was studied by
\citet{wang2011}, who compared EIS observations with simultaneous measurements from the EUNIS
sounding rocket experiment flown on 2007 November 6. For these observations the ratio of EUNIS to
EIS intensity was found to be approximately constant as a function of wavelength, suggesting that
the shape of the EIS pre-flight effective areas were accurate. The mean intensity ratio was 1.22,
indicating a decrease in response of about 20\% since launch. This corresponds to a decay constant
of about 5.6 years.

The long-term evolution of the EIS sensitivity was investigated by \citet{mariska2013}, who the
measured the intensity of several emission lines observed in the quiet Sun from late 2006 to early
2012. He found that the absolute intensities for \ion{Fe}{8} 185.213\,\AA\ and \ion{Si}{7}
275.352\,\AA\ decayed with a time constant of about 20 years during the first 6 years of the
mission, suggesting that the calibration was very stable. This analysis is predicated on the
assumption that intensities measured in the quiet Sun are relatively constant over time.
Intensities in \ion{He}{2} 256.317\,\AA\ showed a much more rapid decline, but this line is
blended with coronal emission and \citet{mariska2013} concluded that the evolution of these
intensities over time was influenced by solar cycle effects.

\citet{delzanna2013} noted that there were several line ratios observed with EIS that did not
agree with previous observations or atomic calculations. The \ion{Fe}{14} 274.203\,\AA\ to
211.316\,\AA\ ratio, for example, is largely insensitive to variations in temperature and
density. The ratio observed with EIS not only disagreed with theory but changed with time,
suggesting that time-dependent modifications to the pre-flight calibration were needed. Through
the systematic analysis of various line ratios \citet{delzanna2013} provided revised effective
area curves for both EIS channels. \citet{delzanna2013} concluded that the degradation was
primarily in the long wavelength channel and that the effective area in the short wavelength
channel remains constant in time.

There are, however, several limitations to using line ratios for calibration purposes. They
establish the relative but not the absolute calibration. Also, the changes to the calibration
suggested by line ratios may not be unique. There is no way to know how to distribute
modifications to the effective areas without additional information. Furthermore, many of the
lines used in such an analysis are weak and are difficult to measure. Modifications to the
calibration based on weak lines may lead to problems with the intensities of the strong lines that
are more generally used for analysis.

Here we extend the previous work on the EIS calibration in two ways. First, we compare full-disk
mosaics constructed by scanning the EIS slot over the Sun with irradiance observation made by the
EUV Variability Experiment (EVE \citealt{woods2012}) on the \textit{SDO} mission. These
comparisons provide a means of establishing the absolute calibration for EIS. Second, we combine
ultra-deep EIS observations from above the limb in the quiet Sun with a simple temperature model
to simultaneously determine the differential emission measure distribution and the time-dependent
changes to the effective areas which best fit all of the available spectral lines. This analysis
takes advantage of the fact that the quiet corona generally has a very narrow temperature
distribution \citep[e.g.,][]{raymond1997,feldman1998,landi2002,warren2009}.

\section{Comparisons With EVE}

\begin{figure*}[t!]
\centerline{\hspace{0.39in}\includegraphics[angle=90,width=0.9\linewidth]{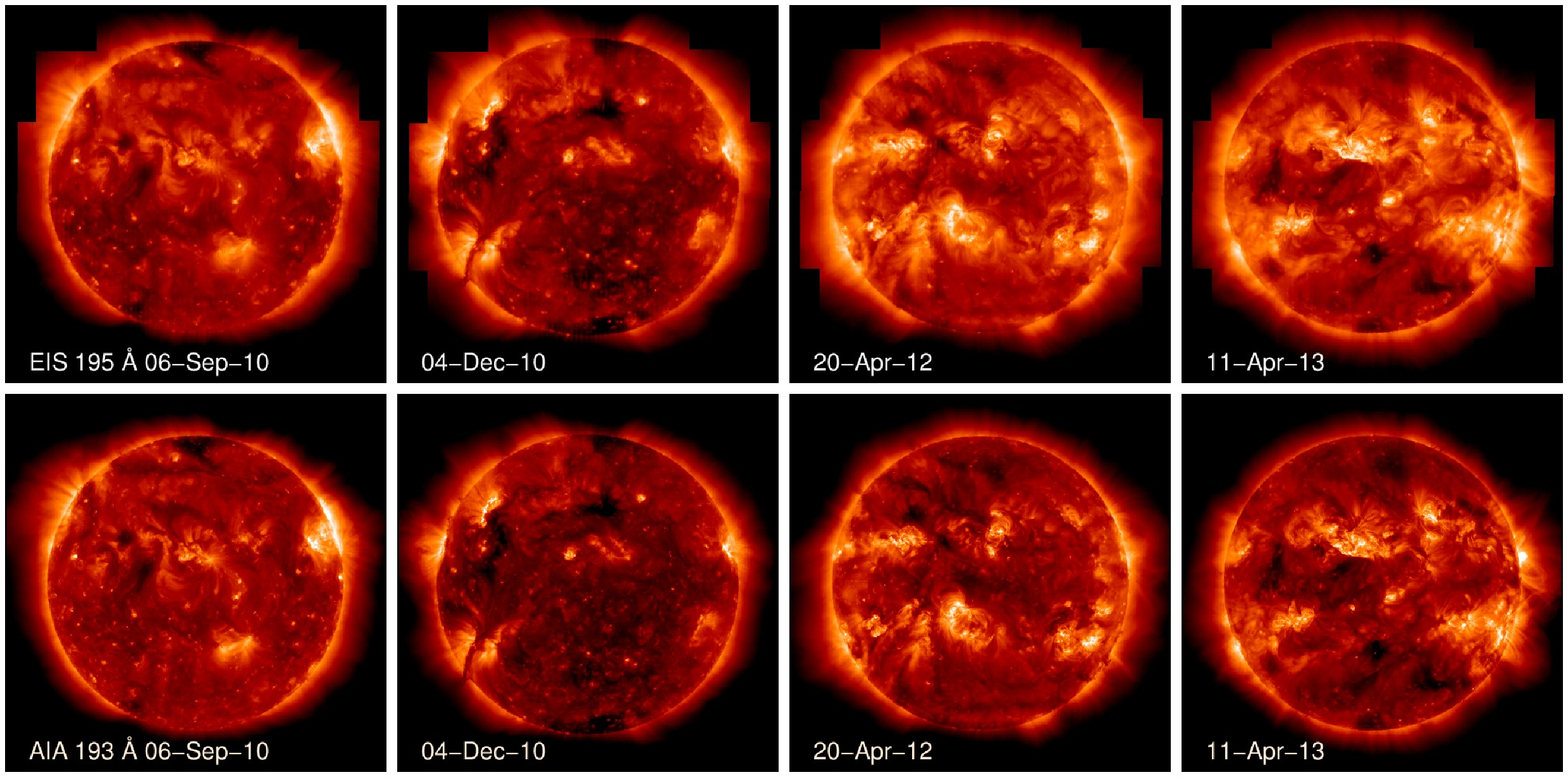}}
\centerline{\includegraphics[angle=90,width=0.9\linewidth]{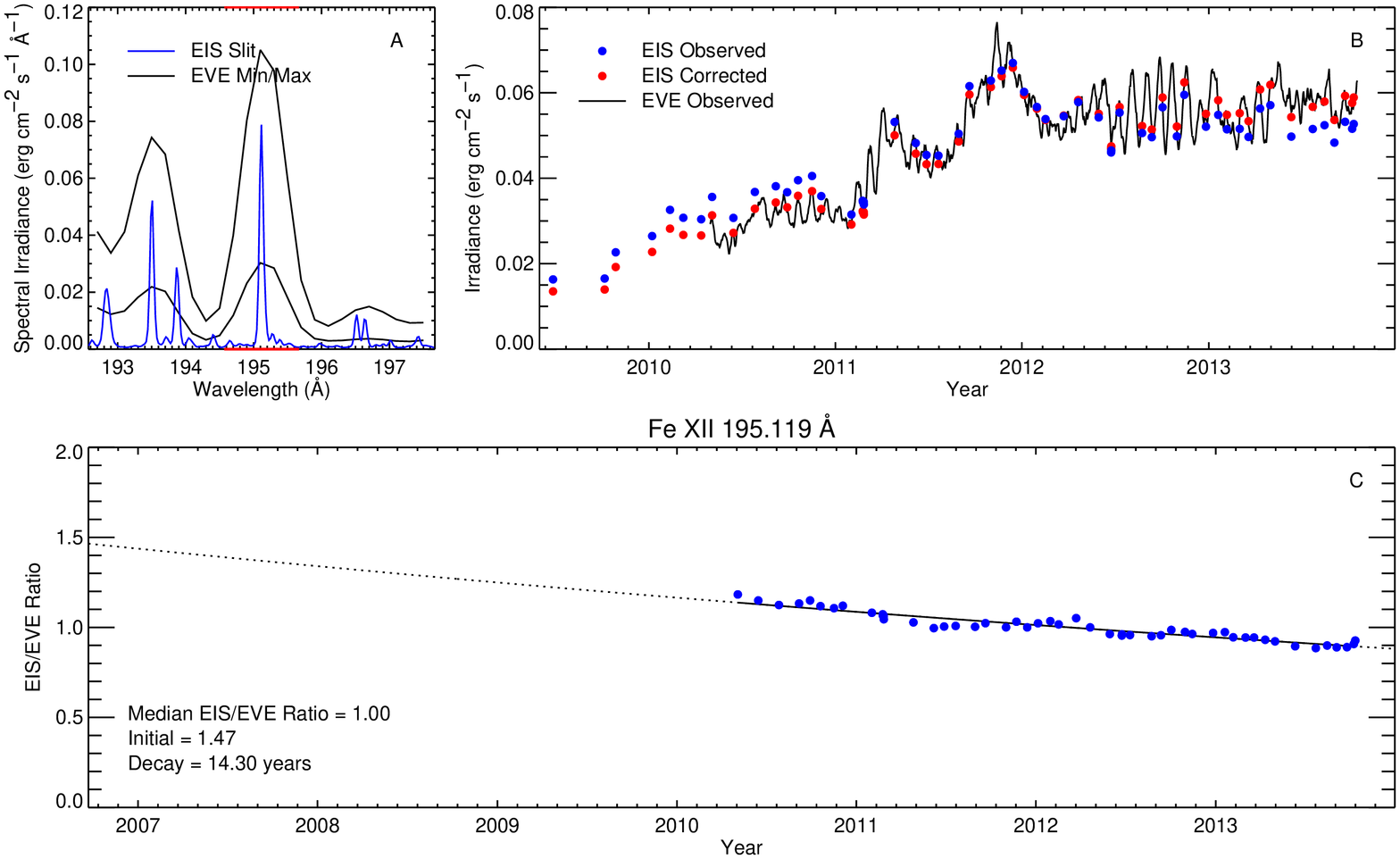}}
\caption{A comparison of irradiances in the \ion{Fe}{12} 195\,\AA\ emission line measured with EIS
  and EVE. The top two rows show representative EIS full-disk mosaics constructed by rastering the
  40\arcsec\ slot as well as AIA 193\,\AA\ images taken the same day. Panel A shows EVE spectral
  irradiances during periods of low and high solar activity. For reference we also show a high
  spectral resolution EIS spectrum from an active region. This spectrum is scaled
  arbitrarily. Panel B shows the EVE and EIS irradiances as a function of time. Panel C shows the
  ratio of EIS to EVE during the SDO mission. The solid line is an exponential fit to the observed
  ratio. The dotted line is an extrapolation of the fit to earlier and later times. EIS
  irradiances corrected for the observed decay are also shown in Panel B.}
\label{fig:eve}
\end{figure*}

EIS has been optimized to provide observations at relatively high spatial and spectral resolution
(see \citealt{korendyke2006} for details). However, in addition to the narrow 1\arcsec\ and
2\arcsec\ slits EIS also can observe with a 40\arcsec\ wide slot. These slot observations are
similar to those obtained from the SO82-A instrument on \textit{Skylab} \citep{tousey1977}. The
40\arcsec\ width of the EIS slot, however, yields images which integrate over only about 1\,\AA\
of the solar spectrum. This limits the amount of overlap between spectral features. To build up a
larger field of view the slot is stepped across the Sun and a short exposure is taken at each
position. By re-pointing the spacecraft a number of times a nearly complete scan of the Sun is
possible within a few hours of observing.

In June of 2009 EIS began executing a full-disk scan of the Sun approximately once every three
weeks. Observations are taken in \ion{Fe}{10} 180.401\,\AA, \ion{Fe}{12} 195.119\,\AA,
\ion{Fe}{13} 202.044 and 203.826\,\AA, \ion{He}{2} 256.317\,\AA, \ion{Fe}{16} 262.984\,\AA,
\ion{Si}{7} 275.368, and \ion{Fe}{15} 284.160\,\AA. Beginning on 2013 March 19 \ion{Fe}{14}
211.316\,\AA\ and 274.203\,\AA, \ion{Si}{10} 258.375\,\AA, and \ion{Si}{10} 264.233\,\AA\ were
added to the observing program.

Each of these observations has been processed to remove the CCD pedestal, dark current, and spikes
from warm and hot pixels. These data are also calibrated using the pre-flight effective areas. The
spatially resolved intensity (or radiance) measured on the Sun is related to the irradiance
measured at earth by
\begin{equation}
  F(\lambda) = \frac{a}{R^2}\sum_{i=1}^N I_i(\lambda),
\end{equation}
where $a$ is the area on the Sun imaged in a detector pixel, $R$ is the earth-Sun distance, and
$N$ is the number of EIS pixels in the full-disk mosaic. Because the EIS mosaic requires several
hours to obtain and the \textit{Hinode} spacecraft pointing often drifts over time, each raster is
cross-correlated with either an EIT 195\,\AA\ \citep{delaboudiniere1995} or an AIA 193\,\AA\
\citep{lemen2012} image to correct for this. Several example EIS full-disk mosaics are shown in
Figure~\ref{fig:eve}. Note that each mosaic covers only a finite portion of the corona above the
limb. Based on comparisons with EIT and AIA images we estimate that EIS underestimates the
irradiance by a few percent. Some EIS images have gaps due to data dropouts and data with more
than 2.5\% missing pixels are not used for our analysis.

For each full-disk mosaic we compare the irradiance inferred from EIS with the corresponding
irradiance measured by EVE. EVE provides absolutely calibrated, spatially unresolved measurements
of the spectral irradiance at a resolution of about 1\,\AA. The absolute calibration of EVE is
monitored using redundant channels and rocket underflights \citep{hock2012}. For comparison with
EIS we read the version 3 daily merged EVE data, convert the spectrum to cgs units, and integrate
over a 1\,\AA\ bandpass centered on the line of interest. The ratio of EIS to EVE irradiances as a
function of time for \ion{Fe}{12} 195.119\,\AA\ is shown in Figure~\ref{fig:eve}. This comparison
suggests that the EIS calibration is changing very slowly over time, generally consistent with the
results of \citet{mariska2013}, although the time constant we measure is 14.5 years instead of
20.2 years. This comparison also suggests that while EIS and EVE agree on an absolute scale during
most of the period from 2010 through 2013, the absolute calibration of EIS at this wavelength is
unlikely to have been correct at launch.

Inspection of the EIS to EVE ratios at the other wavelengths suggests that both the absolute and
relative calibration differs from the pre-flight measurements. A summary of the median EIS/EVE
ratios observed from the beginning of the \textit{SDO} mission to 2013 May 1 are given in
Table~\ref{table:eve}.  In the long wavelength channel the EIS irradiances are generally below
those measured by EVE and show a much more rapid decay (approximately 9 to 12 years). In the short
wavelength channel the EIS to EVE ratio varies from below 1 at the shortest wavelengths to
approximately 1.5 at the longest wavelengths. The time constants are also systematically larger in
the short wavelength channel, ranging from 15 to 27 years. We note that at several of these
wavelengths there is significant blending in the EVE spectra as well as significant overlap in the
EIS slot images. This complicates the interpretation of the ratios for these lines.

As we will discuss in detail later in the paper, the complex behavior in the effective areas as a
function of wavelength and time is broadly consistent with the results of \cite{delzanna2013}. To
fully characterize the EIS calibration, however, more detailed analysis is required. In the next
section we will combine very deep EIS exposures above the limb with a simple temperature model to
infer the best-fit effective areas as a function of time.

\begin{deluxetable}{rcrr}
\tablewidth{2.75in}
\tablecaption{EIS to EVE Irradiance Ratios\tablenotemark{a}}
\tablehead{
\multicolumn{1}{c}{Line} &
\multicolumn{1}{c}{EIS/EVE} &
\multicolumn{1}{c}{$\tau$} &
\multicolumn{1}{c}{Comment}
}
\startdata
       Fe XI 180.401 &       0.63 &       20.6 & \\
     Fe VIII 185.213 &       0.54 &        --- & Blended \\
      Fe XII 195.119 &       1.00 &       14.3 & \\
     Fe XIII 202.044 &       1.27 &       18.7 & \\
     Fe XIII 203.826 &       1.42 &       24.8 & Blended \\
      Fe XIV 211.316 &       1.52 &        --- & \\
       He II 256.317 &       0.49 &       10.3 & Blended \\
        Si X 258.375 &       0.36 &        --- & Blended \\
      Fe XVI 262.984 &       0.36 &       10.8 & \\
         S X 264.233 &       0.34 &        --- & Blended \\
      Fe XIV 274.203 &       0.54 &        --- & \\
      Si VII 275.368 &       0.40 &        4.8 & Blended  \\
       Fe XV 284.160 &       0.72 &       11.0 & 
\enddata
\tablenotetext{a}{The EIS/EVE ratio is the median ratio from the beginning of the \textit{SDO}
  mission to 2013 May 1. $\tau$ is the time constant in years derived from an exponential fit to
  the EIS/EVE ratio. Some EIS lines have been observed only very recently in the full-disk mosaics
  and time constants are not yet available.}
\label{table:eve}
\end{deluxetable}

\section{Quiet Sun Emission Measure}

\begin{figure*}[t!]
\centerline{\includegraphics[angle=90,width=\linewidth]{%
    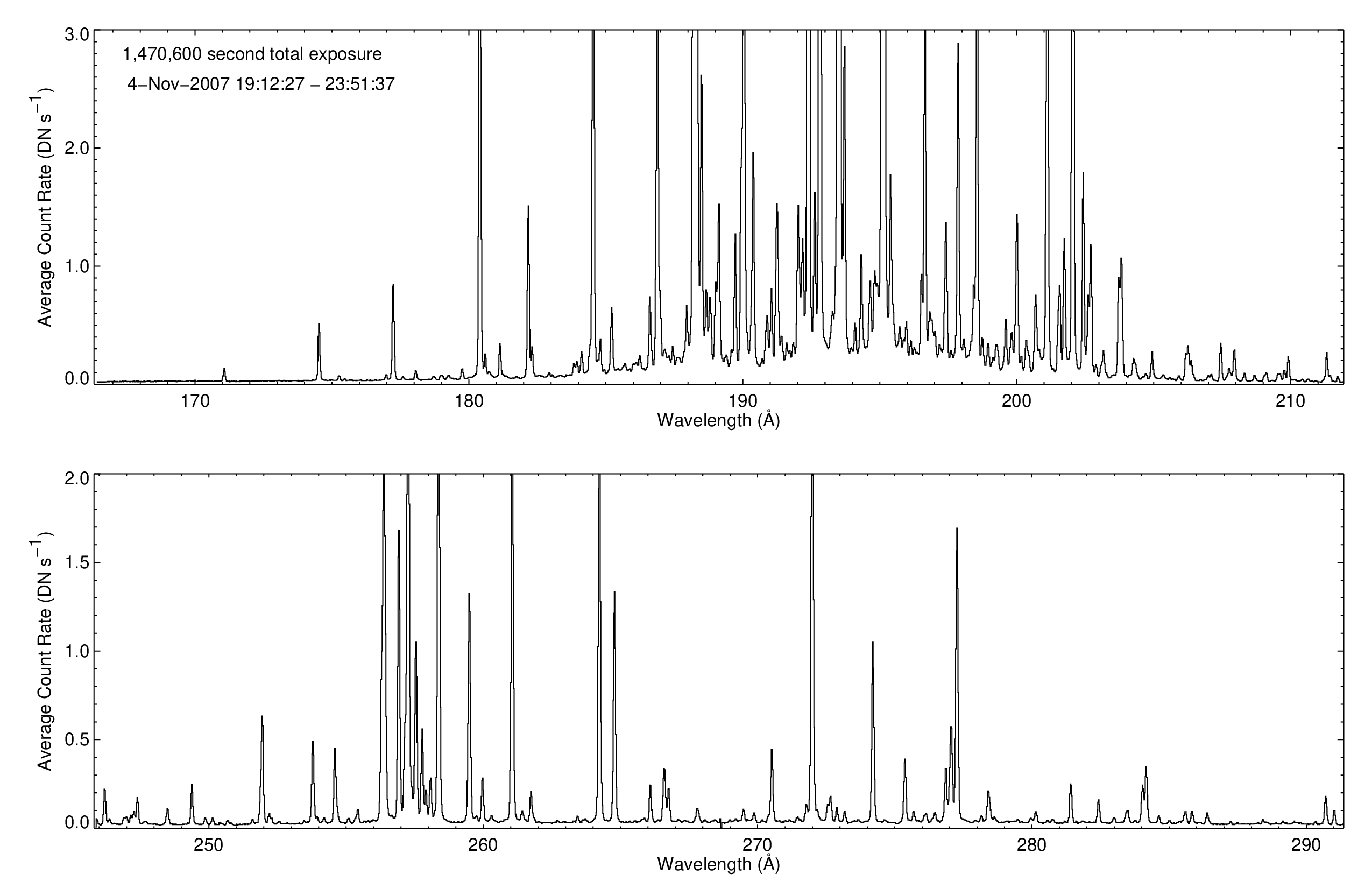}}
\caption{An ultra-deep spectrum of the quiet Sun above the limb constructed by averaging 38 300\,s
  exposures and 129 pixels along the slit for an effective exposure time of 1,470,600\,s. The
  scale in the Y direction has been cutoff to emphasize the weak lines. The peak count rates are
  approximately 50\,DN~s$^{-1}$.}
\label{fig:spec}
\end{figure*}

\begin{figure*}[t!]
\centerline{\includegraphics[angle=90,width=0.95\linewidth]{%
    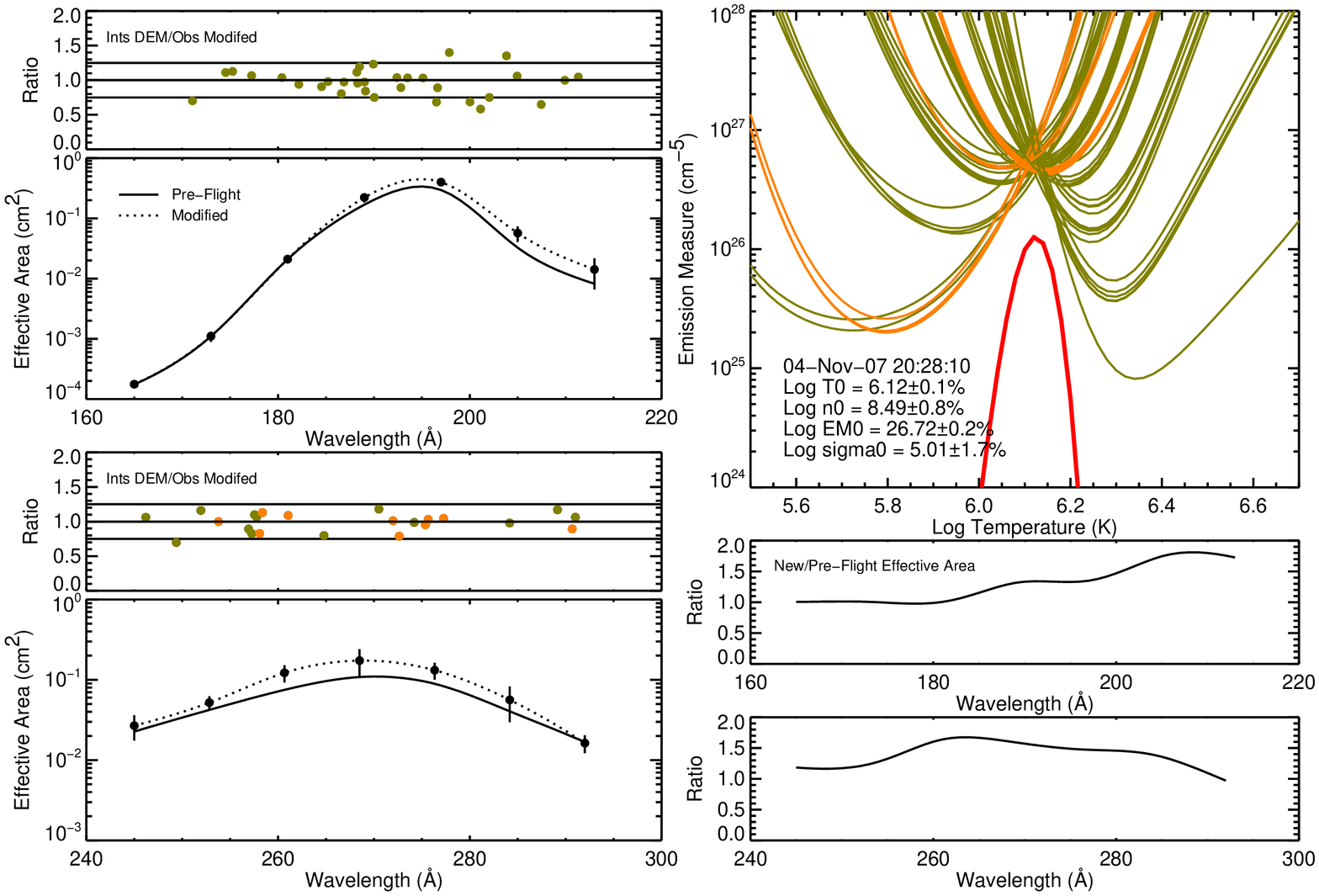}}
\caption{Simultaneous optimization of the differential emission measure and effective areas for
  the spectrum shown in Figure~\ref{fig:spec}. The left panels show the pre-flight and corrected
  effective areas as a function of wavelength for both EIS detectors. The left panels also show
  the ratio of the observed intensities using the corrected effective areas. The panels on the
  lower right show the ratio of the corrected to the pre-flight effective areas. The panel on the
  upper right shows the best-fit differential emission measure distribution and EM-loci curves for
  each line. In these plots green corresponds to Fe lines and orange to Si lines.}
\label{fig:dem}
\end{figure*}

Many previous studies have indicated that the distribution of temperatures above the limb is very
narrow \citep[e.g.,][]{raymond1997,feldman1998,landi2002,warren2009}. These results suggest that a
single Gaussian is a good representation for the differential emission measure, which we write as
\begin{equation}
\xi(T) = \frac{EM_0}{\sigma_T\sqrt{2\pi}}
   \exp\left[-\frac{(T-T_0)^2}{2\sigma_T^2}\right].
\end{equation}
We can convolve the differential emission measure with the plasma emissivity for an emission line
to compute the line intensity using
\begin{equation}
  I_\lambda = \frac{1}{4\pi}\int\epsilon_\lambda(n_e,T)\xi(T)\,dT.
  \label{eq:ints}
\end{equation}
Here the plasma emissivity is the radiated power (erg s$^{-1}$) divided by the square of the
electron number density, $n_e$. For this work we use version 7.1 of the CHIANTI atomic physics
database \citep[e.g.,][]{dere1997,landi2013} to compute the plasma emissivities. To facilitate
rapid calculations we have pre-computed grids of emissivities as a function of temperature and
density. We assume the CHIANTI ionization equilibria and the coronal abundances of
\citet{feldman1992}. Since many of the Fe emission lines observed at these wavelengths have some
density sensitivity, we leave the density as a free-parameter to be determined through the fitting
process and we do not exclude density sensitive lines from our analysis.

For a given intensity the number of photons that are detected in a pixel on the CCD is equal to
\begin{equation}
N_{\rm photons} = \frac{a}{R^2}I_\lambda A(\lambda) t,
\label{eq:cal1}
\end{equation}
where $a$ is the area of the Sun imaged in a pixel and $R$ is the earth-Sun distance, $I_\lambda$
is the intensity at the surface of the Sun in units of photon cm$^{-2}$ s$^{-1}$ sr$^{-1}$,
$A(\lambda)$ is the effective area, and $t$ the exposure time. The effective area includes
information on the collecting area of the mirrors and the efficiency of the various optical
elements, filters, and the detectors \citep{lang2006}. The photons that strike the CCD generate
electrons that are read out by the detector electronics. The number of electrons is equal to
\begin{equation}
N_e = N_{\rm photons}\frac{hc}{\lambda}\frac{1}{3.65\,\textrm{eV}}.
\label{eq:cal2}
\end{equation}
Finally, the detector electronics divide the number of electrons recorded by a gain factor to
produce a ``data number'' value that is sent to the ground in the telemetry stream
\begin{equation}
{\rm DN} = N_e/G.
\label{eq:cal3}
\end{equation}
For EIS the gain is set at 6.3.

For most differential emission measure calculations with EIS the pre-flight effective areas from
\citet{lang2006} have been used. Here we have the more ambitious goal of using the observations to
solve for both the best-fit differential emission measure parameters $\{EM_0, \sigma_T, T_0,
n_e\}$ and effective areas simultaneously. To model the variation in the effective area we
introduce a series of equally spaced spline knots that are used to interpolate the effective areas
as a function of wavelength.  The parameter to be optimized in the calculation is the ratio of the
corrected to pre-flight effective area $\{r_k(\lambda_k,t)\}$. 

For this work we assume 7 knots for each EIS wavelength range.  With a total of 14 knots and 4 DEM
variables there are many free parameters in the model. While it is easy to include enough emission
lines to make the problem well posed, it is important to consider both the temperature of
formation as well as the wavelength for each line. Lines from many different ionization stages are
needed to constrain the DEM while multiple emission lines at each wavelength are needed to
constrain the determination of the effective areas.

\begin{figure*}[t!]
\centerline{\includegraphics[angle=90,width=0.95\linewidth]{%
    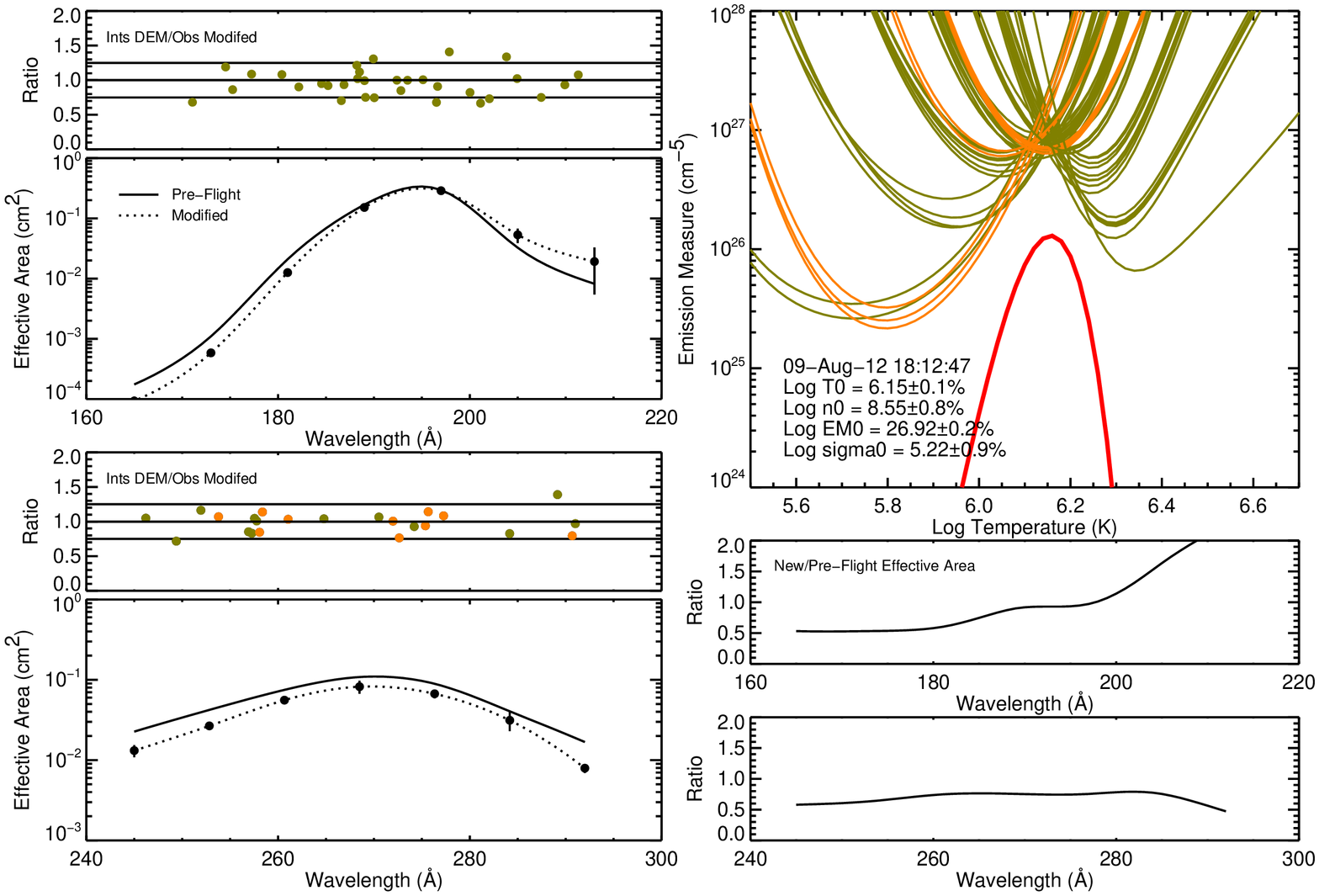}}
\caption{The same as Figure~\ref{fig:dem} but for observations taken later in the mission. The
  effective area in the long wavelength channel is about a factor of 2 lower than the pre-flight
  value.}
\label{fig:dem2}
\end{figure*}

Measuring emission from lines that span the temperature range of the quiet corona is relatively
easy. Emission from lines of \ion{Fe}{8} to \ion{Fe}{16} are often present above the limb. The
challenge for our approach is measuring emission lines over the complete wavelength range of each
detector. This is difficult at the ends of the detectors where the effective areas are very small
and count rates are generally low. To overcome this we have made use of full CCD rasters with very
long exposure times taken by EIS in the quiet Sun above the limb. The EIS study {\tt
  EL\_FULL\_CCD\_W\_SUMER}, which takes 7 300\,s exposures over a narrow
$14\arcsec\times512\arcsec$ area, is particularly useful for this application. We have processed
the data for each of these observations using {\tt eis\_prep} to remove dark current and hot
pixels from the detector exposures. We do not apply the calibration at this stage and leave the
observations in units of data number (DN). We then select an area along the slit and compute the
distribution of DN at each wavelength. We use the median and standard deviation to characterize
the intensity and the uncertainty in the intensity at each wavelength. Using the median, as
opposed to the average, helps limit the impact of residual ``warm pixels'' on the intensity. Some
pixels on the EIS detector produce very high count rates even for exposures taken with the shutter
closed. These ``hot pixels'' are easily identified in dark exposures. Other pixels produce
slightly elevated levels of signal during dark exposures and are more difficult to identify.

In Figure~\ref{fig:spec} we show the average spectrum from on observation of 7 consecutive runs of
{\tt EL\_FULL\_CCD\_W\_SUMER}. The observations began on 2007 November 4 19:12 and ended at on the
same date at 23:51 UT. The EIS field of view was centered at (990\arcsec,-50\arcsec) about
22\arcsec\ above the limb of the Sun. The central 129 pixels along the slit have been averaged
over 38 exposures (11 exposures were corrupted in transmission to the ground) for a total of 4,902
intensity measurements at each wavelength. Since each exposure is 300\,s the spectrum represents
1,470,600\,s of effective exposure time and allows weak lines at the ends of the detector to be
measured.

Using the average spectrum we compute intensities for each emission line of interest by fitting
the profile with a Gaussian and a linear background. For lines that are close in wavelength
multiple Gaussians are fit simultaneously. For this work we focus on emission lines from Fe and Si
to minimize abundance effects. At this stage the intensities are converted from DN to physical
units using the pre-flight effective areas and equations~\ref{eq:cal1}--\ref{eq:cal3}.

Equations \ref{eq:ints} and \ref{eq:cal1} show that the observed count rate is proportional to the
product of the effective area and the emission measure. This means we cannot solve our system of
equations without additional constraints. Our primary assumption is that the EIS effective area
near 195\,\AA\ follows the decay derived from the EIS/EVE comparisons that is shown in
Figure~\ref{fig:eve}. Since observations with the AIA 193\,\AA\ channel are generally consistent
with the absolutely calibrated EVE measurements at the same wavelength \citep{boerner2013}, this
assumption helps insure that any combined EIS-AIA-EVE analysis uses a consistent
inter-calibration.

We further assume that the trend measured during the \textit{SDO} mission can be extrapolated back
to the launch of \textit{Hinode}. While this assumption should be regarded with considerable
skepticism, it is consistent with the simple exponential decays measured by \citet{mariska2013}
from launch to mid 2012.

\begin{deluxetable*}{lrrr@{\,$\pm$\,}rr@{\,$\pm$\,}rrr}
\tablewidth{5.5in}
\tabletypesize{\footnotesize}
\tablecaption{Intensities Measured in the Quiet Corona above the Limb 2007 November 4 with
  EIS\tablenotemark{a}}
\tablehead{
\multicolumn{1}{c}{} &
\multicolumn{6}{c}{Intensity} &
\multicolumn{1}{c}{} \\
[.3ex]\cline{2-8}\\[-1.6ex]
\multicolumn{1}{c}{Line} &
\multicolumn{1}{c}{\textit{DN/s}} &
\multicolumn{1}{c}{\textit{photons/s}} &
\multicolumn{1}{c}{\textit{pre-flight}} &
\multicolumn{1}{c}{$\sigma$} &
\multicolumn{1}{c}{\textit{corrected}} &
\multicolumn{1}{c}{$\sigma$} &
\multicolumn{1}{c}{\textit{DEM}} &
\multicolumn{1}{c}{$R$} 
}
\startdata
Fe IX 171.073        &       0.37 &       0.12 &    1110.50 &     254.98 &    1096.75 &     251.82 &     769.41 &       1.43 \\ 
Fe X 174.532         &       1.72 &       0.56 &     839.70 &     185.35 &     842.92 &     186.06 &     935.89 &       0.90 \\ 
Fe X 175.263         &       0.13 &       0.04 &      50.64 &      14.80 &      51.12 &      14.95 &      57.57 &       0.89 \\ 
Fe X 177.239         &       2.93 &       0.96 &     473.10 &     104.28 &     483.36 &     106.54 &     515.45 &       0.94 \\ 
Fe XI 180.401        &      18.25 &       6.11 &     845.41 &     186.07 &     850.53 &     187.20 &     882.36 &       0.96 \\ 
Fe XI 182.167        &       5.23 &       1.77 &     130.93 &      28.83 &     126.22 &      27.80 &     118.64 &       1.06 \\ 
Fe X 184.536         &      20.61 &       7.05 &     246.86 &      54.34 &     217.93 &      47.98 &     197.35 &       1.10 \\ 
Fe VIII 185.213      &       2.10 &       0.72 &      21.03 &       4.64 &      18.06 &       3.99 &      17.71 &       1.02 \\ 
Fe VIII 186.601      &       2.54 &       0.88 &      18.41 &       4.06 &      14.98 &       3.31 &      12.06 &       1.24 \\ 
Fe XII 186.880       &      13.81 &       4.79 &      94.34 &      20.77 &      76.01 &      16.73 &      73.97 &       1.03 \\ 
Fe XI 188.216        &      92.24 &      32.20 &     488.42 &     107.48 &     377.58 &      83.09 &     421.00 &       0.90 \\ 
Fe XI 188.299        &      64.98 &      22.69 &     339.18 &      74.64 &     261.65 &      57.58 &     250.45 &       1.04 \\ 
Fe IX 188.497        &       8.75 &       3.06 &      44.17 &       9.72 &      33.91 &       7.46 &      40.49 &       0.84 \\ 
Fe XI 189.017        &       2.78 &       0.97 &      12.91 &       2.84 &       9.81 &       2.16 &       9.55 &       1.03 \\ 
Fe XI 189.123        &       5.39 &       1.89 &      24.63 &       5.42 &      18.67 &       4.11 &      15.71 &       1.19 \\ 
Fe IX 189.941        &       6.03 &       2.13 &      24.48 &       5.39 &      18.36 &       4.04 &      22.62 &       0.81 \\ 
Fe X 190.038         &      24.64 &       8.68 &      98.63 &      21.71 &      73.93 &      16.27 &      55.48 &       1.33 \\ 
Fe XII 192.394       &      65.10 &      23.23 &     199.53 &      43.91 &     149.07 &      32.81 &     154.64 &       0.96 \\ 
Fe XI 192.813        &      43.46 &      15.54 &     128.50 &      28.28 &      96.14 &      21.16 &      85.79 &       1.12 \\ 
Fe XII 193.509       &     149.48 &      53.65 &     419.01 &      92.21 &     314.32 &      69.17 &     324.68 &       0.97 \\ 
Fe XII 195.119       &     238.89 &      86.45 &     618.54 &     136.14 &     465.15 &     102.38 &     479.81 &       0.97 \\ 
Fe XIII 196.525      &       2.54 &       0.92 &       6.57 &       1.45 &       4.90 &       1.08 &       3.35 &       1.46 \\ 
Fe XII 196.640       &      11.43 &       4.17 &      29.73 &       6.54 &      22.13 &       4.87 &      19.71 &       1.12 \\ 
Fe IX 197.862        &      10.15 &       3.72 &      29.00 &       6.39 &      21.10 &       4.65 &      29.54 &       0.71 \\ 
Fe XIII 200.021      &       5.80 &       2.15 &      26.21 &       5.77 &      17.82 &       3.93 &      12.22 &       1.46 \\ 
Fe XIII 201.121      &      16.18 &       6.04 &     106.92 &      23.53 &      69.76 &      15.35 &      40.65 &       1.72 \\ 
Fe XIII 202.044      &      35.26 &      13.21 &     335.62 &      73.87 &     211.35 &      46.52 &     158.66 &       1.33 \\ 
Fe XIII 203.826      &       3.96 &       1.50 &      71.01 &      15.66 &      41.98 &       9.26 &      56.75 &       0.74 \\ 
Fe XIII 204.937      &       0.87 &       0.33 &      20.60 &       4.56 &      11.81 &       2.62 &      12.51 &       0.94 \\ 
Fe X 207.449         &       1.19 &       0.46 &      44.75 &       9.91 &      24.76 &       5.49 &      16.07 &       1.54 \\ 
Fe XIII 209.916      &       0.72 &       0.28 &      42.58 &       9.48 &      23.65 &       5.27 &      23.61 &       1.00 \\ 
Fe XIV 211.316       &       0.82 &       0.32 &      60.79 &      13.53 &      34.27 &       7.63 &      35.85 &       0.96 \\ 
Fe XIII 246.208      &       0.81 &       0.37 &      25.72 &       5.71 &      21.95 &       4.87 &      23.30 &       0.94 \\ 
Fe XII 249.388       &       0.83 &       0.38 &      21.53 &       4.79 &      18.45 &       4.11 &      12.87 &       1.43 \\ 
Fe XIII 251.953      &       2.30 &       1.08 &      46.44 &      10.26 &      38.51 &       8.51 &      44.72 &       0.86 \\ 
Si X 253.791         &       1.87 &       0.88 &      31.99 &       7.05 &      25.21 &       5.56 &      25.20 &       1.00 \\ 
Fe XI 256.925        &       6.15 &       2.93 &      82.44 &      18.15 &      57.44 &      12.65 &      51.31 &       1.12 \\ 
Fe X 257.262         &      15.59 &       7.44 &     204.15 &      44.95 &     140.28 &      30.89 &     115.84 &       1.21 \\ 
Fe XI 257.547        &       4.18 &       2.00 &      53.66 &      11.81 &      36.45 &       8.03 &      40.01 &       0.91 \\ 
Fe XI 257.772        &       1.85 &       0.88 &      23.35 &       5.15 &      15.72 &       3.47 &      16.77 &       0.94 \\ 
Si IX 258.073        &       0.81 &       0.39 &       9.97 &       2.22 &       6.63 &       1.47 &       5.50 &       1.21 \\ 
Si X 258.375         &      14.18 &       6.79 &     171.97 &      37.85 &     113.13 &      24.90 &     127.88 &       0.88 \\ 
Si X 261.058         &      10.05 &       4.87 &     102.61 &      22.58 &      62.49 &      13.75 &      68.00 &       0.92 \\ 
Fe XIV 264.787       &       3.59 &       1.77 &      29.82 &       6.57 &      17.88 &       3.94 &      14.27 &       1.25 \\ 
Fe XIV 270.519       &       1.70 &       0.85 &      12.02 &       2.65 &       7.71 &       1.70 &       9.10 &       0.85 \\ 
Si X 271.990         &      12.12 &       6.11 &      86.83 &      19.10 &      56.66 &      12.47 &      57.31 &       0.99 \\ 
Si VII 272.641       &       0.59 &       0.30 &       4.32 &       0.97 &       2.84 &       0.63 &       2.24 &       1.27 \\ 
Fe XIV 274.203       &       3.99 &       2.03 &      30.77 &       6.77 &      20.54 &       4.52 &      20.27 &       1.01 \\ 
Si VII 275.368       &       1.47 &       0.75 &      12.08 &       2.66 &       8.14 &       1.80 &       7.77 &       1.05 \\ 
Si VII 275.665       &       0.22 &       0.11 &       1.84 &       0.42 &       1.24 &       0.28 &       1.28 &       0.97 \\ 
Si X 277.265         &       6.99 &       3.60 &      66.13 &      14.55 &      44.99 &       9.90 &      47.06 &       0.96 \\ 
Fe XV 284.160        &       1.29 &       0.68 &      24.87 &       5.50 &      17.92 &       3.96 &      17.56 &       1.02 \\ 
Fe XIV 289.151       &       0.05 &       0.03 &       1.67 &       0.81 &       1.46 &       0.71 &       1.71 &       0.85 \\ 
Si IX 290.687        &       0.67 &       0.36 &      26.92 &       6.03 &      25.60 &       5.74 &      22.87 &       1.12 \\ 
Fe XII 291.010       &       0.32 &       0.17 &      13.40 &       3.10 &      12.99 &       3.00 &      13.81 &       0.94
\enddata
\tablenotetext{a}{The intensities computed from the pre-flight and corrected effective areas as
  well as the intensity computed from the DEM are in units of erg cm$^{-2}$ s$^{-1}$ sr$^{-1}$ the
  other intensities are in units of data number per second and detected photons per second. $R$ is
  the ratio of corrected to DEM calculated intensity.}
\label{table:ints1}
\end{deluxetable*}

Unfortunately there is only a single EIS line near both 171 and 284\,\AA\ so we introduce an
additional constraint that the ratio of the corrected to pre-flight effective area is the same as
it is at the next wavelength. That is, we assume that $r_k = r_{k+1}$ for the first knot in the
short wavelength band.

Finally, with the very long exposure times for these observations \ion{Fe}{12} 193.509 and
195.119\,\AA\ are often saturated. We replace any measured intensities for these lines with
intensities inferred from the intensity of \ion{Fe}{12} 192.394\,\AA\ and the ratios from CHIANTI
of 2.1 and 3.1, respectively.

With these constraints we use the {\tt MPFIT} package \citep{markwardt2009} for implementing
Levenberg-Marquardt least-squares minimization. We use $\{10^{27}, 10^{5.5}, 10^{6.2}, 10^9\}$ as
initial conditions for the emission measure parameters and, except for the knot closest to
195\,\AA, we assume ${r_k = 1}$ for the initial effective area ratios. The value for the knot
closest to 195\,\AA\ is fixed to the value of the EIS/EVE ratio shown in Figure~\ref{fig:eve}. At
each iteration new intensities are computed from the emission measure and effective area ratio
parameters. Iteration continues until the differences between the computed and observed
intensities reaches a minimum.  In Figure~\ref{fig:dem} we show the solution for the 2007 November
4 spectrum discussed earlier. A summary of the intensities is given in
Table~\ref{table:ints1}. For this date the EIS/EVE decay curve indicates that the effective area
at 195\,\AA\ is 1.35 times the pre-flight value. The best-fit solution indicates that the
effective areas need to be raised at wavelengths above 195\,\AA\ as well as in the central region
of the long wavelength detector.

\begin{figure*}[t!]
\centerline{\includegraphics[width=0.95\linewidth]{%
    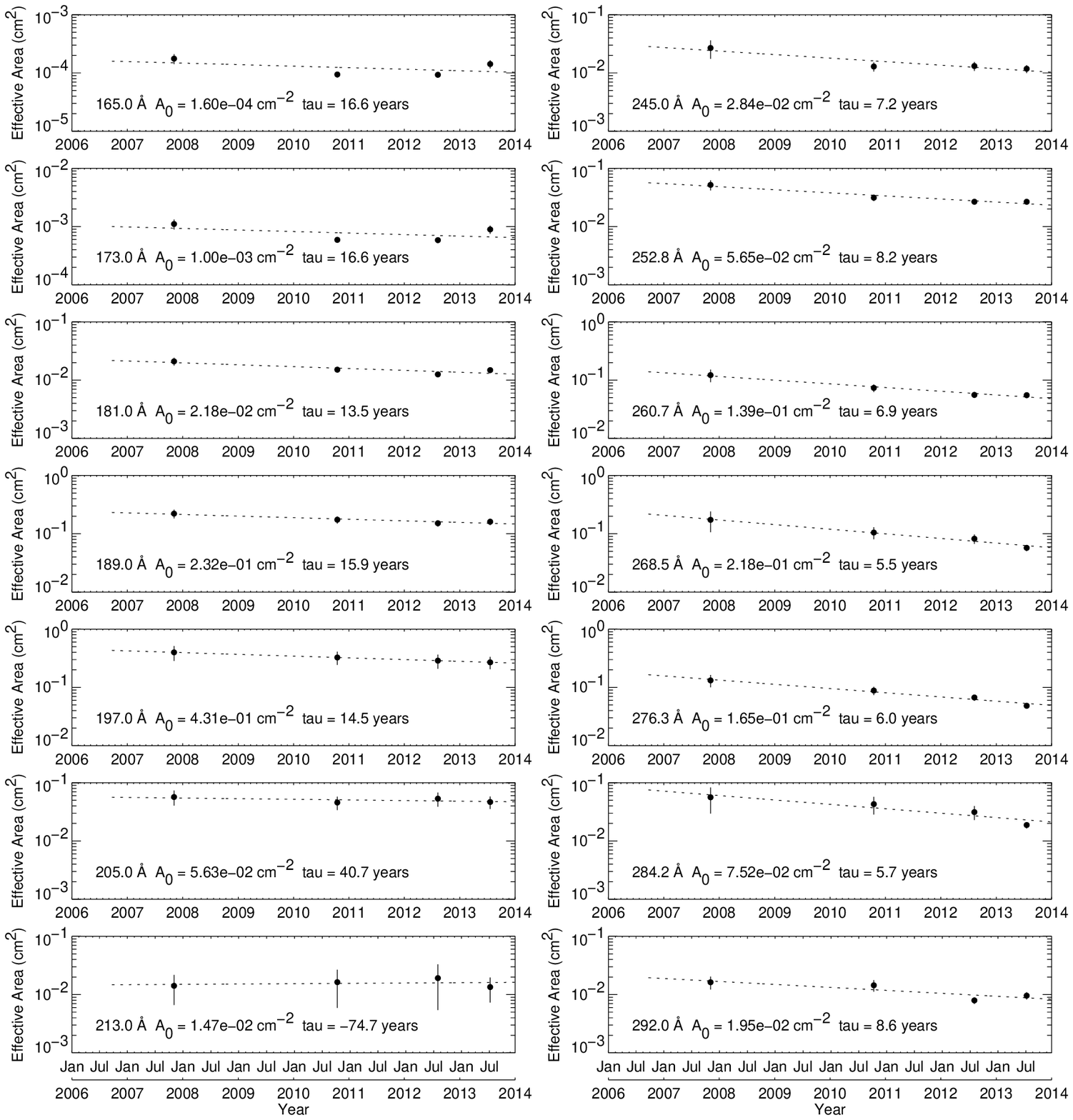}}
\caption{The evolution of the effective areas at the spline knots. Error bars derived from the
  least-squares fitting are shown. The dotted lines are exponential fits to the effective area as
  a function of time.}
\label{fig:ea}
\end{figure*}

For comparison we have also run this calculation with the effective area ratios fixed at unity so
that the pre-flight calibration is used in computing the DEM.  The parameters derived for the DEM
are very similar to those computed with the corrected effective area curves, but now systematic
trends in the calculated intensities are evident. The computed intensities for lines at
wavelengths above 195\,\AA\ and near 260\,\AA\ are generally too small while the intensities at
180\,\AA\ and below are generally too large.

\begin{figure*}[t!]
\centerline{\includegraphics[width=\linewidth]{%
    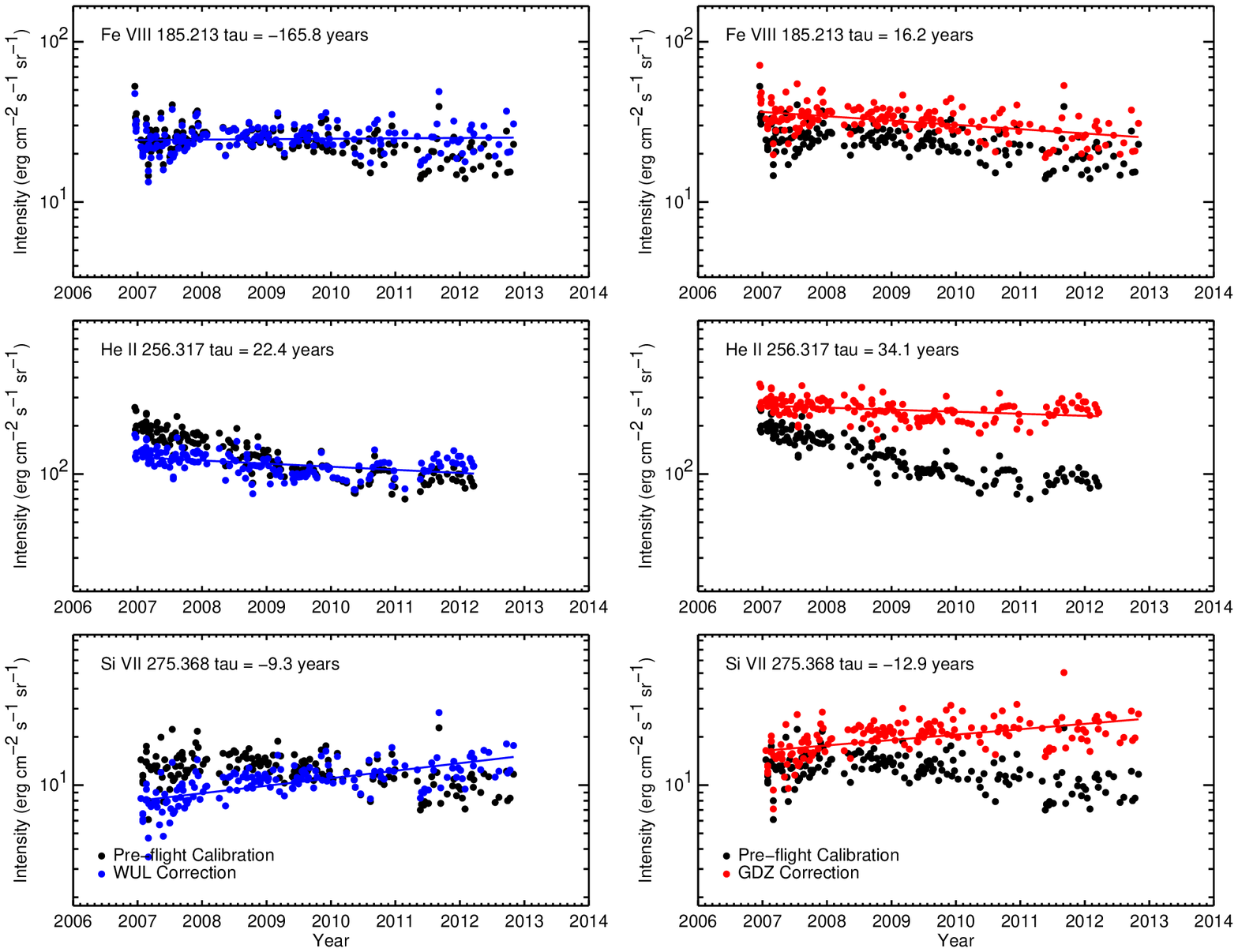}}
\caption{EIS synoptic observations re-calibrated using the effective areas derived in this paper
  as well as those from \citet{delzanna2013}. These data are from \citet{mariska2013}. The new
  effective area curves can account for the secular declines in the \ion{He}{2} 256.317\,\AA\ and
  \ion{Fe}{8} 185.213\,\AA\ intensities, but overcorrect \ion{Si}{7} 275.368\,\AA.}
\label{fig:synop}
\end{figure*}

The application of this analysis to the 2007 November 4 data yields new effective areas curves for
one time. In Figure~\ref{fig:dem2} we show the analysis for off-limb observations taken 2012
August 9, almost 5 years after the data shown in Figure~\ref{fig:dem}. The resulting effective
areas curves are similar in shape to those calculated for the earlier data, but the magnitude has
diminished at many wavelengths.

There are, unfortunately, only a few observations taken above the limb in the quiet corona which
have the required signal to noise for this analysis. We have only been able to perform the DEM
calculations for a total 5 data sets, two of which were taken the same day.  In
Figure~\ref{fig:ea} we plot our effective area as a function of time for several selected
wavelengths. Also shown are exponential fits of the form
\begin{equation}
 A_{\rm corrected}(\lambda,t) = A_{\rm corrected}(\lambda,t_0)
 \exp\left[-\frac{t-t_0}{\tau(\lambda)}\right],
\label{eq:ea}
\end{equation}
where $t_0$ is chosen to be 2006 September 22, the \textit{Hinode} launch date, and
$\tau(\lambda)$ is the wavelength dependent time constant. An exponential model is consistent with
time evolution of the sensitivity found by \citet{mariska2013}. The decay in the sensitivity of
the long wavelength channel is fairly uniform, with $\tau\approx$ 6--7 years. The decay in the
sensitivity of the short wavelength channel is slower and less uniform, with $\tau\approx$ 9 years
at the shortest wavelengths and a slight increase of sensitivity at the end of the short
wavelength detector.

Our analysis is generally consistent with the results of \citet{delzanna2013}. Both sets of
corrections indicate a reduction of the EIS effective area at the shortest wavelengths in the short
wavelength channel as well as a relative increase near the central part of the long wavelength
detector (see Figure~8 from \citealt{delzanna2013}). Our analysis, however, also suggests a relative
increase in the effective area at the longest wavelengths in the short wavelength detector. Our
comparisons with EVE indicate a slow decrease in the sensitivity of the short wavelength detector,
which we have incorporated into our analysis. Using observation of the transition region
intensities taken in the quiet Sun, \citealt{delzanna2013} concluded that sensitivity in the short
wavelength detector has remained approximately constant during the mission.

\section{Comparisons}

In this section we investigate the application of the time-dependent effective areas to various
datasets that have been used to characterize the EIS calibration. We have written a routine that
implements Equation~\ref{eq:ea} by evaluating the time constants at the wavelengths of the spline
knots and interpolating to the wavelength of interest. We can then recalibrate an intensity
derived from the pre-flight calibration using
\begin{equation}
I_{\rm corrected}(\lambda,t) = I_{\rm pre\_flight}(\lambda)
      \frac{A_{\rm pre\_flight}(\lambda)}{\rm A_{corrected}(\lambda,t)}.
\end{equation}

In Figure~\ref{fig:synop} we show the intensities from the EIS synoptic program which periodically
measures the intensities of selected emission lines in the quiet Sun. These data were analyzed by
\citet{mariska2013}. The revised effective area curves correct the observed secular declines in
both the \ion{Fe}{8} 185.213\,\AA\ and \ion{He}{2} 256.317\,\AA\ lines. Our corrections, however,
introduce a secular increase in the \ion{Si}{7} 275.368\,\AA. The origin of this increase is
unclear. 

\begin{figure}[t!]
\centerline{\includegraphics[width=\linewidth]{%
    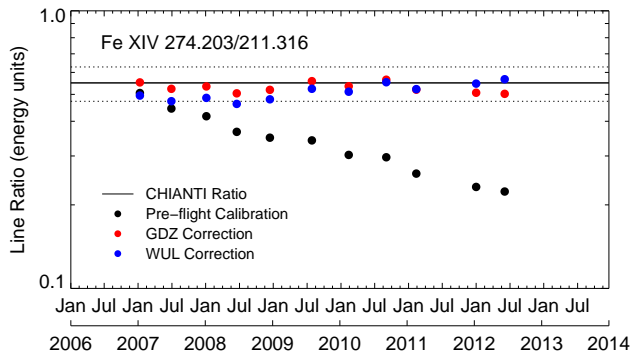}}
\caption{The evolution of the EIS \ion{Fe}{14} 274.203/211.316\,\AA\ ratio as a function of
  time. These measurements are taken from active region observations. The solid line represents
  the theoretical value. The dotted lines represent the uncertainties in the theoretical ratio
  assuming a statistical uncertainty of 10\% for each intensity measurement. }
\label{fig:fe14}
\end{figure}

The problems with EIS observations of the \ion{Fe}{14} 274.203/211.316\,\AA\ ratio were noted
first by \citet{delzanna2013}. As shown in Figure~\ref{fig:fe14}, the observed ratio, which should
be approximately constant, decays exponentially during the \textit{Hinode} mission.  The corrected
effected areas derived from our analysis account for much of the deviation from the theoretical
value. Note that our analysis suggests that the change in the ratio is due to a combination of a
decrease in sensitivity at 274\,\AA\ and a slight increase in sensitivity at 211\,\AA. The
corrections suggested by \citet{delzanna2013} are even closer to theory. These corrections account
for the change in the ratio using only a decrease in the sensitivity at 274\,\AA.

\begin{figure*}[t!]
\centerline{\includegraphics[width=2.4in]{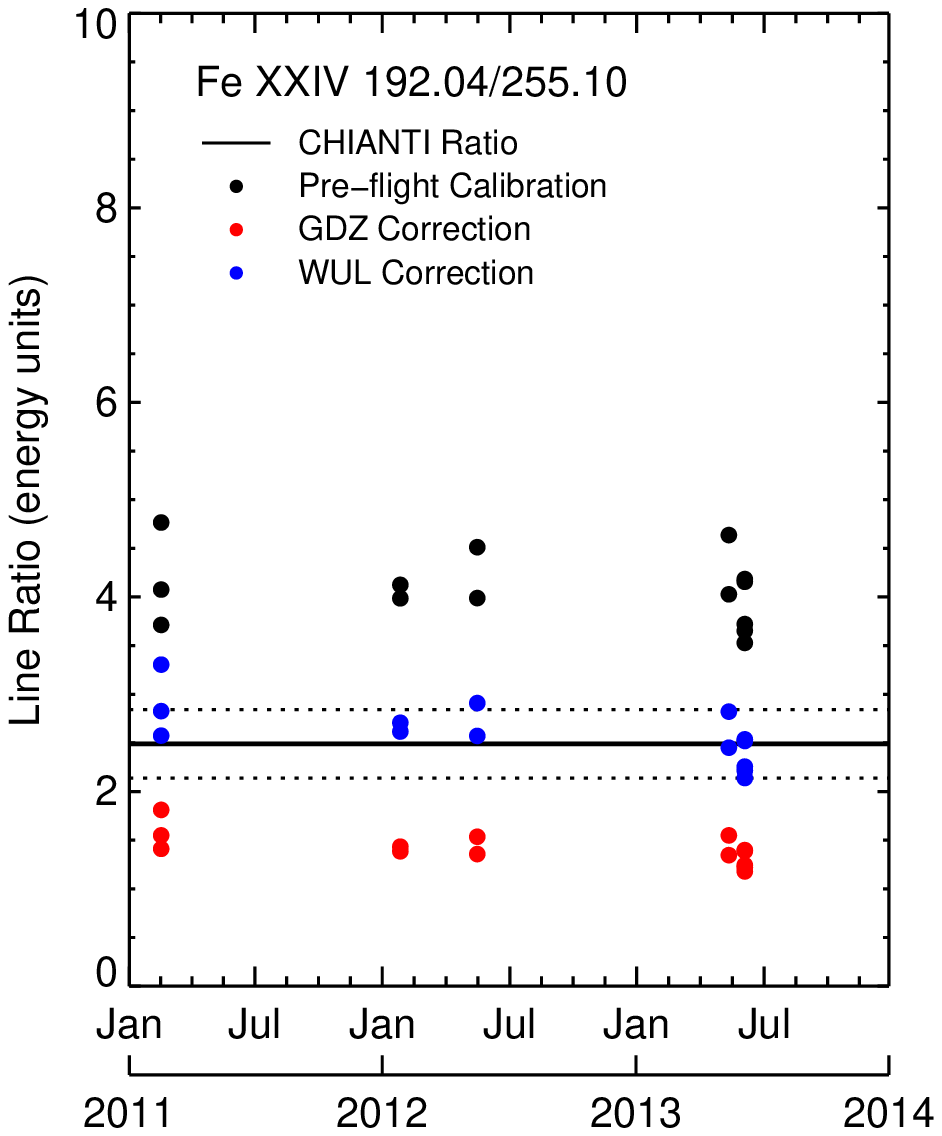}
            \includegraphics[width=4.6in]{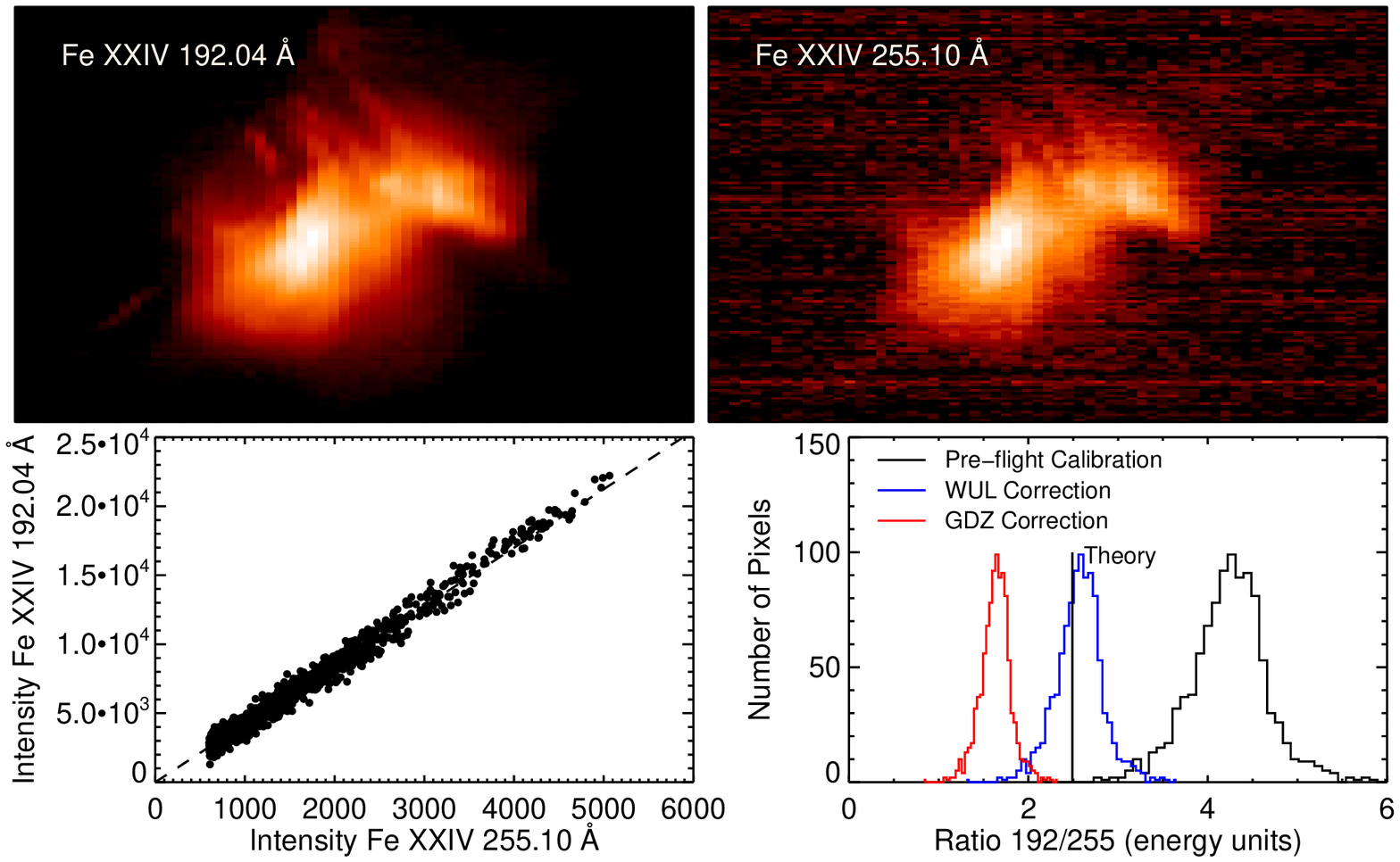}}
          \caption{Observed and corrected \ion{Fe}{24} 192.04/255.10\,\AA\ ratios as a function of
            time. For a temperature of 10\,MK the theoretical ratio is approximately 2.49 in
            energy units. Recent observations are close to 4.5 and show considerable scatter. The
            left panels show ratios derived from several spot checks of flare observations. The
            right panels show more systematic observations from a new observing sequence that
            provides higher spatial resolution. These data from an M5.9 flare that was observed by
            EIS on 2013 June 7. These rasters are from a scan that began at about 23:13 UT. The
            scatter plot shows the intensities for each line as well as a constant ratio of 4.24,
            which the median value for these observations. Only intensities above a threshold in
            the 255\,\AA\ line are used. The histograms show the distributions of observed and
            corrected ratios. The corrected effective areas derived from our analysis bring the
            observations into good agreement with theory.}
\label{fig:fe24}
\end{figure*}

\citet{young2013} and \citet{delzanna2013} noted that the \ion{Fe}{24} 192.04/255.10\,\AA\ ratio
also disagreed with theory. In Figure~\ref{fig:fe24} we show observed ratios for these lines
derived from several observations taken over the past several years. The line ratio observed with
EIS is typically about 4, much different than the well established theoretical ratio of 2.49 (in
energy units).

Recently EIS has observed a large flare above the limb using an autonomous observing mode. These
observations provide relatively high spatial resolution rasters during much of the event and
extensive measurements of the \ion{Fe}{24} line ratio. The analysis of these observations indicate
that the corrected effective areas derived from our analysis can bring the observed ratio into
agreement with theory. The corrections suggested by \citet{delzanna2013} overcorrect the ratio
somewhat.

Early in 2013 \ion{Fe}{14} 211.316\,\AA\ and 274.203\,\AA\ were added to the EIS full-disk mosaic
program. As indicated in Table~\ref{table:eve}, for the \ion{Fe}{14} 211.316\,\AA\ mosaics we find
an EIS/EVE ratio of about 1.5. This is consistent with our finding that the pre-flight effective
areas at this wavelength are too small. We measure an EIS/EVE irradiance ratio of 0.55 for
\ion{Fe}{14} 274.203\,\AA. Again, this is consistent with our finding that the effective area at
this wavelength is currently below the pre-flight value. Similarly, the EIS/EVE irradiance ratios
for \ion{Fe}{9} 180.401\,\AA, \ion{Fe}{13} 202.044\,\AA, \ion{He}{2} 256.317\,\AA, and
\ion{Fe}{15} 284.160\,\AA\ are all consistent with the general trends in the effective areas
derived from our analysis. We note, however, that the time constants derived from the EIS/EVE
irradiance ratios are not always consistent with those derived from our DEM analysis. For example,
the DEM analysis suggests a time constant of 10.2 years near 180\,\AA\ while the EIS/EVE ratio
indicates 18.0 years.

\section{Summary and Discussion}

We have presented an analysis of the absolute calibration of the EIS instrument on
\textit{Hinode}. This analysis, which attempts to reconcile EIS observations with both EVE
irradiance measurements and the atomic data for most of the available emission lines, confirms
that the changes to the EIS calibration are a complex function of both time and wavelength as
found by \citet{delzanna2013}. 

The corrections proposed here can account for many of the problems with the use of the pre-flight
EIS calibration, such as the changes in the \ion{Fe}{14} 274.203/211.316\,\AA\ and \ion{Fe}{24}
255.10/192.04\,\AA\ line ratios. Ultimately, however, we find that it is not possible to fully
reconcile all of the observations. Some of the time constants inferred from the EIS/EVE ratios,
for example, are different than those inferred from the DEM analysis.  It seems likely that the
wavelength-dependent corrections to the EVE observations are not yet full understood, but this is
unlikely to resolve all of the differences that we have found. The corrections that we propose to
the effective areas imply a secular increase in the \ion{Si}{7} 275.368\,\AA\ quiet Sun
intensities. The origin of this is unclear. One possibility is that we have not fully accounted
for the inhomogeneity in the evolution of the effective area.

\begin{figure}[t!]
\centerline{\includegraphics[width=\linewidth]{%
    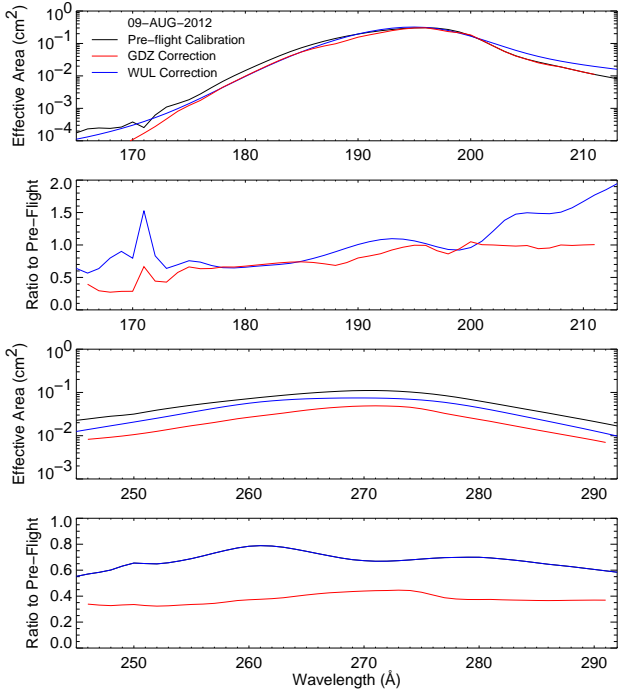}}
\caption{A comparison of the EIS effective areas determined from the pre-flight calibration,
  \cite{delzanna2013}, and this work (WUL). Both the absolute effective areas and the ratios of
  the revised effective areas to the pre-flight values are shown. These curves are for 2012 August
  9, a period for which there is good agreement among all three sets of effective areas near
  195\,\AA. }
\label{fig:gdz}
\end{figure}

\begin{figure}[t!]
\centerline{\includegraphics[width=\linewidth]{%
    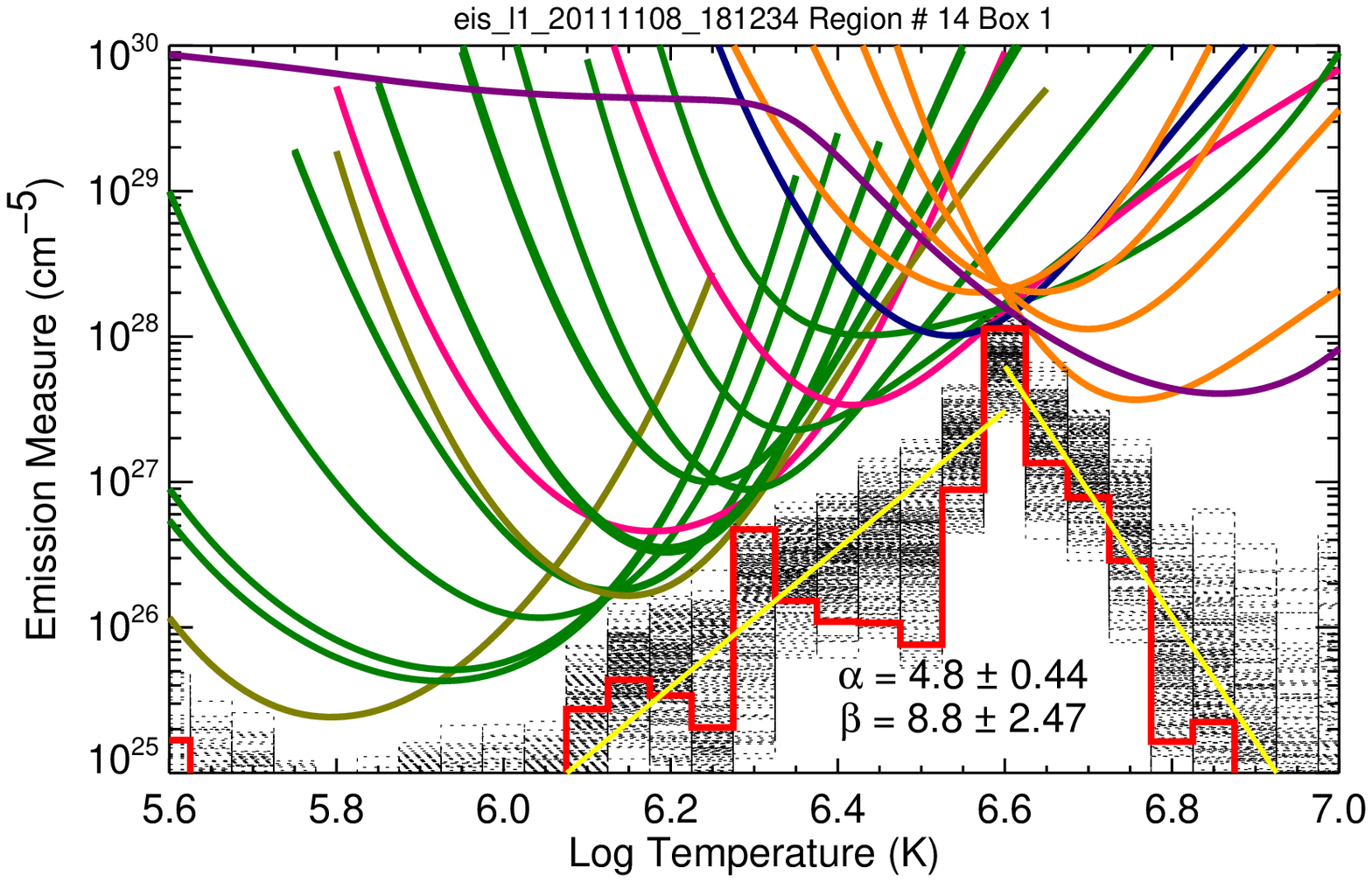}}
\centerline{\includegraphics[width=\linewidth]{%
    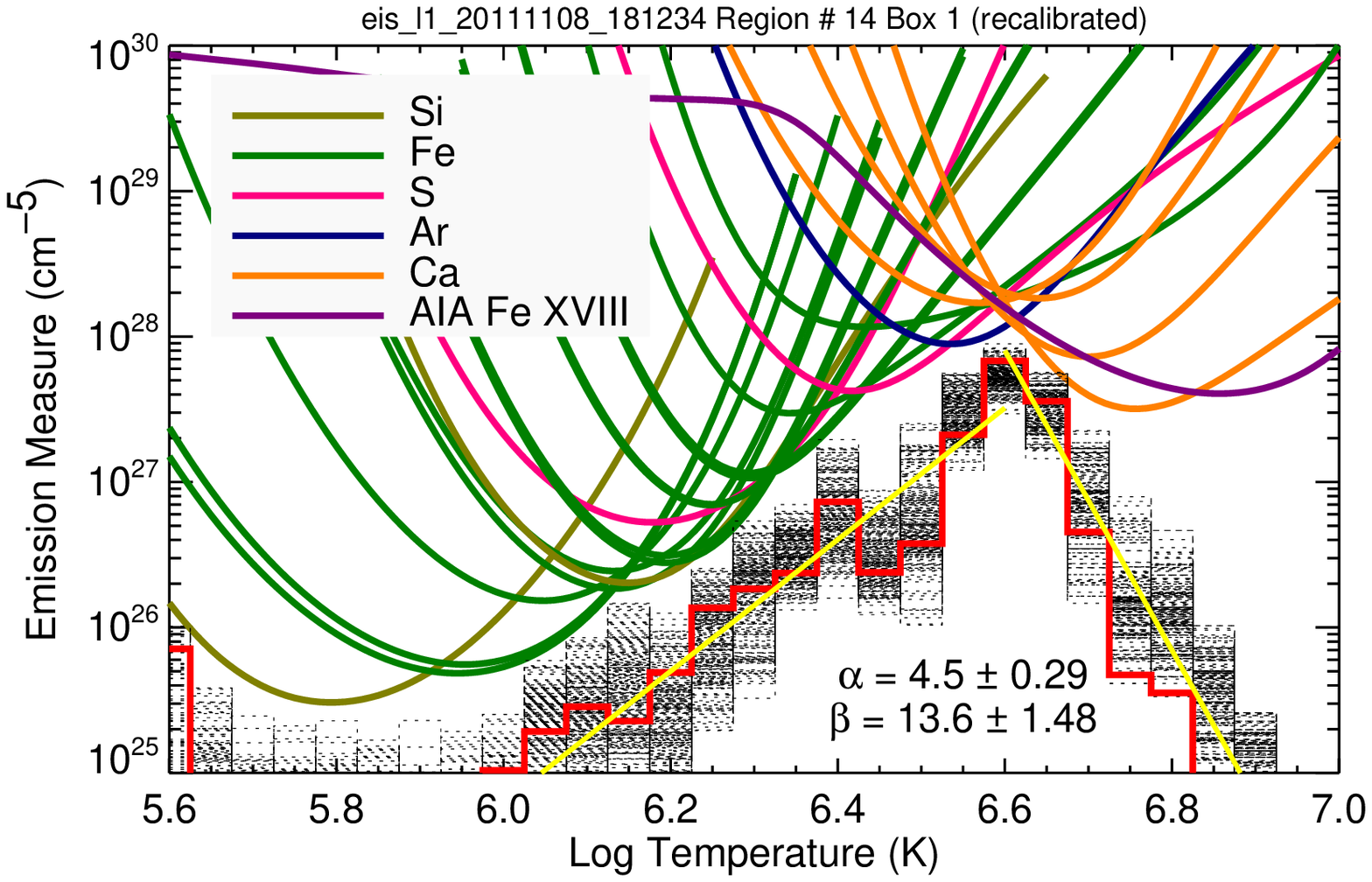}}
\caption{Active region differential emission measure curves derived using the pre-flight (top
  panel) and corrected effective areas (bottom panel). The emission measure distribution is the
  red curve. The black curves are solutions that use perturbed intensities.  The general features
  of the emission measure distribution are largely unchanged.}
\label{fig:dem_eis}
\end{figure}

As is illustrated in Figure~\ref{fig:gdz}, there is good agreement between our results and the
corrections suggested by \cite{delzanna2013} at many wavelengths. There are, however, three
primary differences: 1) We extrapolate the exponential decay observed in the EIS-EVE ratio back to
the beginning of the \textit{Hinode} mission, which suggests higher effective areas at launch. 2)
We find that the effective area above about 200\,\AA\ is higher than the pre-flight value by as
much as 50\%. 3) Our values for the effective areas in the long wavelength channel are generally
about 50\% higher than those given by \cite{delzanna2013}.  The extrapolation of the EIS-EVE trend
is difficult to justify. It does, however, successfully account for the evolution of the
\ion{Fe}{8} 185.213\,\AA\ quiet Sun intensities (see Figure~\ref{fig:synop}). Also, as indicated
in Table~\ref{table:eve}, EIS irradiances 211\,\AA\ implied by the pre-flight calibration are
about 1.52 times those observed with EVE while EIS irradiances at 275\,\AA\ are about 0.54 times
those observed with EVE. Both values are approximately consistent with the effective areas we have
derived through our analysis. At other wavelengths, however, the EIS-EVE ratios are not consistent
with our analysis. These lines are generally strongly blended in EVE and these ratios are more
difficult to interpret. More extensive analysis is required to resolve these issues.

We also cannot reconcile our analysis of the EIS calibration with the EIS-EUNIS inter-calibration.
\citet{wang2011} found a ratio of EIS to EUNIS of approximately 1.22. This ratio was independent
of wavelength, suggesting that the shape of the EIS pre-flight effective areas were correct but
that the instrumental sensitivity had decayed somewhat since launch. However, if we adjust the EIS
intensities found in Table~4 of their paper based on the corrections that we have derived here,
the ratio of EUNIS to EIS jumps to approximately 2 for many wavelengths. Since we tied the
absolute EIS calibration to EVE this implies that the ENUIS absolute calibration is not consistent
with that of EVE. EUNIS was flown again on 2013 April 23 so we will have an opportunity to revisit
this issue.

\begin{deluxetable}{lrrrcrrr}
\tabletypesize{\small}
\tablecaption{An Example Re-Analysis of Active Region Observations\tablenotemark{a}}
\tablehead{
  \multicolumn{1}{c}{}           & 
  \multicolumn{3}{c}{Pre-flight} & 
  \multicolumn{3}{c}{Corrected}  \\ 
  [.3ex]\cline{2-4}\cline{6-8}\\[-1.6ex]
  \multicolumn{1}{c}{Line}      &
  \multicolumn{1}{c}{$I_{obs}$} &
  \multicolumn{1}{c}{$I_{dem}$} &
  \multicolumn{1}{c}{$R$}       &
  \multicolumn{1}{c}{}          & 
  \multicolumn{1}{c}{$I_{obs}$} &
  \multicolumn{1}{c}{$I_{dem}$} &
  \multicolumn{1}{c}{$R$}
}
\startdata
      \ion{Si}{7} 275.368 &      8.8 &      8.4 &     1.0 &  &     11.2 &     12.0 &      0.9 \\
      \ion{Fe}{9} 188.497 &     10.5 &     11.5 &     0.9 &  &     10.8 &     11.7 &      0.9 \\
      \ion{Fe}{9} 197.862 &      8.0 &      7.6 &     1.1 &  &      8.1 &      7.9 &      1.0 \\
     \ion{Fe}{10} 184.536 &     56.5 &     40.8 &     1.4 &  &     73.1 &     48.6 &      1.5 \\
     \ion{Fe}{11} 180.401 &    279.7 &    255.4 &     1.1 &  &    398.8 &    309.4 &      1.3 \\
     \ion{Fe}{11} 188.216 &    139.5 &    124.5 &     1.1 &  &    146.6 &    150.0 &      1.0 \\
      \ion{S}{10} 264.233 &     21.1 &     21.4 &     1.0 &  &     24.5 &     23.3 &      1.1 \\
     \ion{Si}{10} 258.375 &     55.0 &     79.3 &     0.7 &  &     65.4 &     87.4 &      0.7 \\
     \ion{Fe}{12} 192.394 &    128.1 &    122.1 &     1.0 &  &    111.7 &    129.3 &      0.9 \\
     \ion{Fe}{12} 195.119 &    433.1 &    377.4 &     1.1 &  &    388.2 &    399.8 &      1.0 \\
     \ion{Fe}{13} 202.044 &    576.9 &    331.6 &     1.7 &  &    456.4 &    362.7 &      1.3 \\
     \ion{Fe}{13} 203.826 &    657.9 &    381.2 &     1.7 &  &    438.2 &    347.5 &      1.3 \\
     \ion{Fe}{14} 264.787 &    330.4 &    376.7 &     0.9 &  &    388.3 &    391.8 &      1.0 \\
     \ion{Fe}{14} 270.519 &    185.6 &    210.5 &     0.9 &  &    241.7 &    226.0 &      1.1 \\
     \ion{Fe}{15} 284.160 &   4921.1 &   6169.3 &     0.8 &  &   6481.2 &   6575.5 &      1.0 \\
      \ion{S}{13} 256.686 &    380.7 &    455.0 &     0.8 &  &    476.7 &    486.4 &      1.0 \\
     \ion{Fe}{16} 262.984 &    728.8 &    732.3 &     1.0 &  &    832.2 &    684.3 &      1.2 \\
     \ion{Ar}{14} 194.396 &     50.7 &     53.4 &     0.9 &  &     44.5 &     48.6 &      0.9 \\
     \ion{Ca}{14} 193.874 &    282.7 &    197.8 &     1.4 &  &    245.8 &    174.5 &      1.4 \\
     \ion{Ca}{15} 200.972 &    247.3 &    170.3 &     1.5 &  &    225.5 &    144.2 &      1.6 \\
     \ion{Ca}{16} 208.604 &    127.1 &    107.9 &     1.2 &  &     82.1 &     87.2 &      0.9 \\
     \ion{Ca}{17} 192.858 &    110.8 &    138.6 &     0.8 &  &     96.0 &    104.6 &      0.9 \\
                   AIA 94 &     14.4 &     17.6 &     0.8 &  &     14.4 &     13.9 &      1.0
\enddata
\tablenotetext{a}{The observed intensities ($I_{obs}$) are from region \#14 in
  \protect{\citet{warren2012}}. Calculated intensities ($I_{dem}$) are from the an emission
  measure inversion. The variable $R$ is $I_{obs}/I_{dem}$. EIS intensities are in units of erg
  cm$^{-2}$ s$^{-1}$ sr$^{-1}$. AIA intensities are in units of DN s$^{-1}$.}
\label{table:ints2}
\end{deluxetable}

The time constants for changes in the EIS calibration that we find are generally long relative to
the length of the mission. While this implies that the calibration is changing relatively slowly,
it does amplify the uncertainties in the revised effective areas. The sparsity of the available
data also significantly limits our ability to discern more complicated patterns in the evolution
of the instrumental sensitivity. For example, it is possible that the sensitivity has decayed at
different rates at different times during the mission, but our analysis cannot account for this.

We have written software routines that return both pre-flight and modified effective area curves
for a given wavelength and observing time. These routines have been incorporated into the
SolarSoftware distribution of the standard EIS analysis software. One strategy for working with
EIS observations is to continue using the pre-flight calibration during the initial data
processing and then correct the intensities to account for new effective areas. This would allow
for different corrections to be considered. 

Given the complex nature of the changes to the calibration it is difficult to assess their impact
on previous analysis. Any analysis that is based on a only few lines is likely to be the most
sensitive to any changes in the effective areas and should be reconsidered. In contrast, any
analysis than involves many lines is likely to remain largely unchanged. To illustrate this we
have recomputed the active region differential emission measures from \citet{warren2012} using our
new effective area curves. In all cases the re-calibrated intensities yield emission measure
distributions that are similar to those derived using the pre-flight calibration. Example DEMs are
shown in Figure~\ref{fig:dem_eis} and the corresponding intensities are given in
Table~\ref{table:ints2}. We note that the inversions computed with the revised calibration
generally have a $\chi^2$ that is about a factor of 2 smaller than those computed with the
pre-flight values for the effective areas.

The presence of time and wavelength dependent changes to the EIS calibration indicate that
comprehensive synoptic observations are an important component of sensitivity monitoring. Simply
measuring several wavelengths in the quiet Sun is insufficient. Our analysis suggests that routine
monitoring of the quiet corona above the limb could be useful for identifying relative changes to
the effective area of spectroscopic instruments over time.  Since such an analysis is likely to
require very deep exposures it would need to be implemented as part of a special observing
campaign. Ultimately, however, new techniques need to be developed that provide simple yet
accurate measurements of sensitivity over time.

%% ------------------------------------------------------------------------------------------
%% --- ACKNOWLEDGMENTS ----------------------------------------------------------------------
%% ------------------------------------------------------------------------------------------

\acknowledgments Hinode is a Japanese mission developed and launched by ISAS/JAXA, with NAOJ as
domestic partner and NASA and STFC (UK) as international partners. It is operated by these
agencies in co-operation with ESA and NSC (Norway). CHIANTI is a collaborative project involving
George Mason University, the University of Michigan (USA) and the University of Cambridge
(UK). The authors would like to thank John Mariska for providing his EIS synoptic time series and
Peter Young for providing the \ion{Fe}{14} intensities. The authors would also like to thank John
Mariska and Peter Young for many enlightening discussions. This work was funded by NASA's
\textit{Hinode} project.

%% ------------------------------------------------------------------------------------------
%% --- REFERENCES ---------------------------------------------------------------------------
%% ------------------------------------------------------------------------------------------

\end{document}